\documentclass[10pt,twocolumn,letterpaper]{article}

\usepackage[pagenumbers]{cvpr} 

\usepackage{algorithm}
\usepackage{textcomp}
\usepackage{amsmath}
\usepackage{multirow}
\usepackage{caption}
\usepackage{booktabs}
\usepackage[noend]{algpseudocode}
\usepackage[dvipsnames]{xcolor}

\definecolor{cvprblue}{rgb}{0.21,0.49,0.74}
\usepackage[pagebackref,breaklinks,colorlinks,citecolor=cvprblue]{hyperref}
\usepackage{blindtext}
\usepackage{graphicx}
\usepackage{caption}
\usepackage{pifont}
\newcommand{\cmark}{\ding{51}}%
\newcommand{\xmark}{\ding{55}}%

\def\paperID{9842} 
\def\confName{CVPR}
\def\confYear{2024}

\title{POPDG: Popular 3D Dance Generation with PopDanceSet}

\author{Zhenye Luo\thanks{equalcontrib.} \and Min Ren \footnotemark[1]\and Xuecai Hu\thanks{Corresponding author.}\and Yongzhen Huang\footnotemark[2]\and Li Yao \\
School of Artificial Intelligence, Beijing Normal University\\
{\tt\small luozy2021@mail.bnu.edu.cn, \{renmin, huxc1208, huangyongzhen, yaoli\}@bnu.edu.cn}
\vspace{-10mm}
}

\begin{document}
\twocolumn[{%
\renewcommand\twocolumn[1][]{#1}%
\maketitle
\begin{center}
\centering
\includegraphics[width=0.9\textwidth,height=4cm]{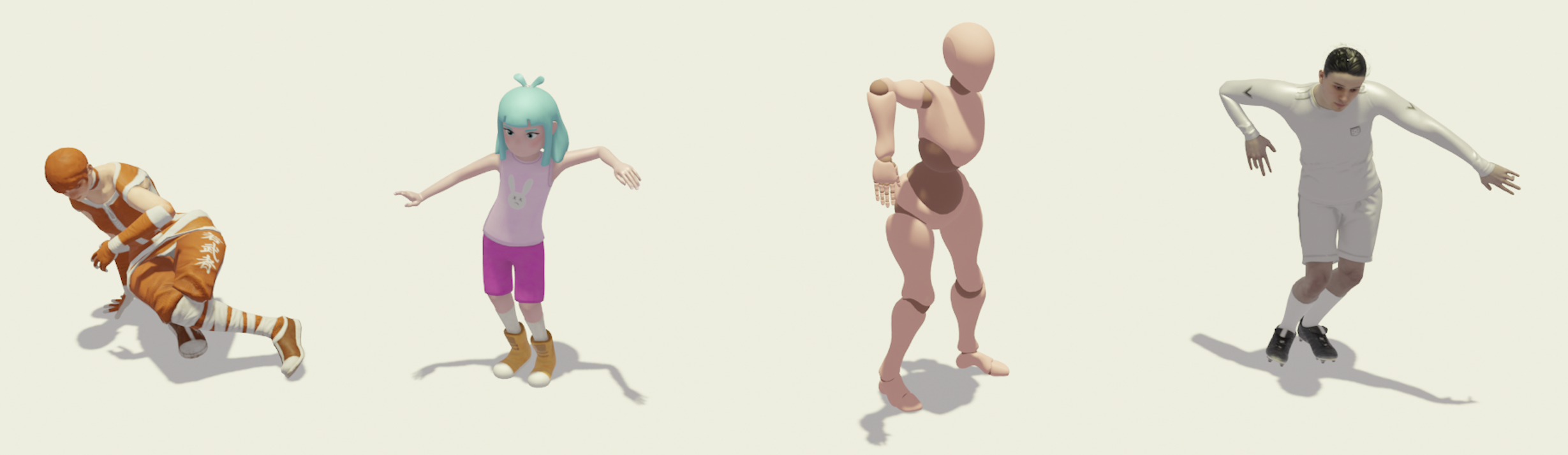}
\captionof{figure}{POPDG, in combination with PopDanceSet, could generate a variety of aesthetically driven popular dances.}
\end{center}%
}]
\footnotetext[1]{These authors contributed equally.}
\footnotetext[2]{Corresponding author.}
\begin{abstract}
Generating dances that are both lifelike and well-aligned with music continues to be a challenging task in the cross-modal domain. This paper introduces PopDanceSet, the first dataset tailored to the preferences of young audiences, enabling the generation of aesthetically oriented dances. And it surpasses the AIST++ dataset in music genre diversity and the intricacy and depth of dance movements. Moreover, the proposed POPDG model within the iDDPM framework enhances dance diversity and, through the Space Augmentation Algorithm, strengthens spatial physical connections between human body joints, ensuring that increased diversity does not compromise generation quality. A streamlined Alignment Module is also designed to improve the temporal alignment between dance and music. Extensive experiments show that POPDG achieves SOTA results on two datasets. Furthermore, the paper also expands on current evaluation metrics. The dataset and code are available at \url{https://github.com/Luke-Luo1/POPDG}.
\end{abstract}    
\section{Introduction}
\label{sec:intro}

Dance is a fundamental artistic expression with a rich history in humanity. Throughout time, humans have utilized dance to convey messages and express emotions \cite{lamothe2019dancing}.The task of music-driven dance generation not only helps choreographers improve the efficiency of creating innovative dances but also facilitates performances by virtual characters. It even extends to the field of neuroscience, assisting researchers in exploring the relationship between human movement and music \cite{brown2008neuroscience}.

This task has long been hampered by the scarcity of publicly available datasets and the limitations in generative model capabilities. As of now, the AIST++ dataset\cite{li2021ai} is among the few with a significant volume of data that is publicly accessible. Despite significant advancements in dance generation models in recent years, issues such as the complexity of training steps, instability in generation, and lack of diversity still exist. This paper introduces the PopDanceSet and the POPDG, aimed at enhancing both the dataset and the model aspects of dance generation.

The AIST++ dataset's limitations include a lack of aesthetically oriented dances, a narrow range of dance and music genres, among others. The dances in this dataset are confined to 10 subcategories of street dance, which hardly encompass the vast array of dance styles in reality. The PopDanceSet, created through a popularity function designed in this paper, filters dance videos that align with popular aesthetics. It represents a significant breakthrough in terms of aesthetically oriented content, diversity in dance types, music genres, and dance movements.

Previous generative models primarily focused on the temporal alignment of music and dance, but even when considering the spatial constraints of dance, they did not delve into the physical interconnections between specific joints. Instead, they approached the task holistically or attempted to learn specific movement patterns \cite{siyao2022bailando}. The alignment between dance and music is also crucial. Prior methods either underestimated this issue or complicated the training process \cite{wu2021dual,wu2021music,siyao2022bailando}. Undoubtedly, these issues impact the overall quality and diversity of generated dances.

This paper specifically proposes a space augmentation algorithm based on Attention Mechanism, forming a dance decoder block to strengthen the spatial connections among joints in dance movements. Furthermore, a streamlined alignment module is designed to encode the spatiotemporal features of music alongside dance, thereby significantly enhancing their rhythmical synchronization.

Finally, in the task of music-driven dance generation, existing evaluation metrics have certain limitations. This paper also proposes evaluation metrics that are suited to this task, thereby enabling a more reasonable assessment of the generated dances.
In summary, the contributions of this paper can be enumerated as follows:
\begin{itemize}
    \item We build the PopDanceSet, reflecting contemporary aesthetic preferences. It significantly enriches the diversity and quantity of dances and music, increases the complexity of dance movements, and offers excellent extensibility for continuous supplementation.
    \item We introduce POPDG(Popular 3D Dance Generation), which is based on iDDPM and achieves a balance between generation quality and diversity. The model pays particular attention to the spatial features of the dancer's body joints, especially proposing the Space Augmentation Algorithm. In addition, our newly designed Alignment Module integrates the spatiotemporal features of music and dance, strengthening the alignment between dance and music.
    \item Extensive experiments were conducted in this study. It was observed that the POPDG produced the exciting results, both on AIST++ and PopDanceSet. And we also make a reasonable extension to the evaluation metrics, making the assessment of dance generation more comprehensive and objective.
\end{itemize}

\begin{table*}
  \centering
  \begin{tabular}{@{}lccccccccc@{}}
    \toprule
    Dataset & Lyrics & Aesthetics & \text{3D Joint}\textsubscript{pos} & \text{3D Joint}\textsubscript{rot} & 2D Kpt & Genres & Subjects & Seconds\\
    \midrule
    Dance with Melody\cite{tang2018dance} & \xmark & \xmark & \cmark & \xmark & \xmark & 4 & - & 5640\\
    GrooveNet\cite{alemi2017groovenet} & \xmark & \xmark & \cmark & \xmark & \xmark & 1 & 1 &  1380 \\
    DanceNet\cite{zhuang2022music2dance} & \xmark & \xmark & \cmark & \xmark & \xmark & 2 & 2 & 3472\\
    EA-MUD\cite{sun2020deepdance} & \xmark & \xmark & \cmark & \xmark & \xmark & 4 & - & 1254\\
    AIST++\cite{li2021ai} & \xmark & \xmark & \cmark & \cmark & \cmark & 10 & 30 & \textbf{18694}\\
    \midrule
    \textbf{PopDanceSet}\textbf{(Ours)} & \textbf{\cmark} & \textbf{\cmark} & \cmark & \cmark & \cmark & \textbf{19} & \textbf{132} & 12819\\
    \bottomrule
  \end{tabular}
  \caption{\textbf{3D Dance Datasets Comparison.} PopDanceSet stands out for its aesthetically oriented content and inclusion of music with corresponding lyrics. Encompassing a broad range of 19 genres and 132 subjects, it offers high diversity over 12,819 seconds of data, establishing itself as a valuable dataset for dance generation research.}
  \label{tab:dataset}
\end{table*}
\section{Related Works}
\label{sec:related works}
\subsection{Music-Dance Dataset}
High-quality dance generation relies on comprehensive and diverse music-dance datasets. Earlier research primarily utilized motion capture technology for limited dataset collection, as seen in \cite{alemi2017groovenet,tang2018dance,zhuang2022music2dance,sun2020deepdance}, or leveraged pose estimation models \cite{cao2017realtime,fang2022alphapose,sengupta2020mm} to derive 2D/3D poses from online dance videos. However, due to the complexities in dance motion capture and the constraints of earlier pose estimation models, these datasets were limited in dance variety, duration, and motion capture quality. A significant advancement was made with AIST++ \cite{li2021ai}, an extension of AIST \cite{tsuchida2019aist}, offering longer durations, precise 3D joint annotations, and high-quality dance movements, setting a new standard in the field. Despite its wide usage, later databases like PMSD\cite{valle2021transflower}, PhantomDance\cite{li2022danceformer}, and MMD\cite{chen2021choreomaster} provide only incremental advancements, mainly providing additional data for specific research tasks without much broader impact due to limited public availability.

\subsection{Human dance generation}
Initially, music-driven dance generation, an autoregressive task, explored the music-dance relationship using traditional machine learning algorithms \cite{fan2011example,lee2013music,ofli2011learn2dance}, but these methods produced dances with limited duration, diversity, and poor adaptability to various melodies and rhythms. The advent of deep learning saw researchers \cite{alemi2017groovenet,tang2018dance,lee2019dancing,lamothe2019dancing,huang2020dance,ye2020choreonet,sun2020deepdance,ren2020self,wu2021dual,wu2021music,li2021autodance,duan2021automatic,chen2021choreomaster,zheng2021adversarial,zhuang2022music2dance} employing CNNs, LSTMs, MLPs, and GCNs to better capture deep features. Despite improved feature extraction and generalizability, generating dances with high diversity remains challenging. With FACT's introduction \cite{li2021ai}, Transformers have gained prominence for their superior temporal feature modeling \cite{kim2022brand,huang2022genre,valle2021transflower}. Further advancements by Bailando\cite{siyao2022bailando,siyao2023bailando++} and EDGE\cite{tseng2023edge} using VQ-VAE, GPT, reinforcement learning, and DDPM have enhanced dance quality and diversity but at the cost of increased training complexity. The stability and overall quality of long-sequence dance generation continue to need enhancement.

\subsection{Diffusion Models}
Diffusion models \cite{ho2020denoising}, a novel class of deep generative models, learn data distributions through reverse denoising processes. They have recently shown superior generative capabilities in image generation, outperforming benchmarks in general tasks \cite{nichol2021improved,rombach2022high}. Additionally, their adaptability in conditional generation tasks makes them highly versatile. Dhariwal\cite{dhariwal2021diffusion} and Ho\cite{ho2022classifier} demonstrated their effectiveness with guided image generation, optimizing the diversity-fidelity trade-off. Their impressive performance extends to various fields, including 3D monocular pose estimation\cite{gong2023diffpose} and text-driven motion generation\cite{zhang2022motiondiffuse}. While closely related to human pose and motion generation with emerging applications in music-driven dance generation \cite{tseng2023edge}, the high standards for quality and diversity in this domain mean diffusion models still necessitate further exploration.


\section{PopDanceSet}
\label{sec:dataset}

\subsection{Popularity Function and Dataset Construction}

Our aim in building PopDanceSet was to address the issues mentioned in section 2.1 while also catering to the aesthetic preferences of contemporary youth. To this end, we developed a popularity function to filter suitable dance videos. We selected BiliBili\cite{bilibili}, the video platform most popular among young people in China, as our data source. Using multiple linear regression and Student's t test \cite{eberly2007multiple}, we identified the variables that influence video popularity, formulated the popularity function, and detailed the verification process in supplementary \cref{sec:popdataset}.
\begin{equation}
      Pop = WN^{T}+b,
  \label{eq:heat function}
\end{equation}
In \cref{eq:heat function}, we define N as \([n_{favorites},n_{danmucounts},\\ n_{views}, n_{likes}, n_{shares}]\), where each term represents the number of favorites, danmu(live comments that scroll over the video, offering an interactive and communal viewing experience) counts, views, likes, and shares respectively. These are weighted by the coefficient vector \(W = [0.0251, 0.0095, 0.8033, 0.0967, 0.0243]\). Additionally, the bias term \(b\) is set to 0.0443. We establish a Pop threshold of 0.85 for selection criteria. Recognizing the inherent advantage of authors with a larger following, we consider only those videos where the view count exceeds the number of followers of the creator, denoted as \(n_{views} > n_{followers}\). Moreover, we opt to exclude videos with frequent changes in camera angles or excessive shaking, to ensure data consistency and quality.

\subsection{Dataset Description}

We collected a total of 263 dance videos, containing 180 pieces of music. In recent years, monocular 3D joint detection technology based on SMPL\cite{loper2023smpl,zhang2022mixste} has made significant progress, providing high-quality detection results. We employed the HybrIK model \cite{li2021hybrik,li2023hybrik} to extract the 3D joint features of the dancers in all videos. Each frame of data has the following parameters:
\begin{itemize}
    \item 24 SMPL pose parameters along with the global scaling, translation and pred\_scores;
\end{itemize}
\begin{itemize}
    \item Predicted camera parameters along with root and translation;
\end{itemize}
\begin{itemize}
    \item 17 COCO-format\cite{ruggero2017benchmarking} human joint locations in 3D;
\end{itemize}
The comparison between PoPDanceSet and other datasets is in \cref{tab:dataset}. It is readily apparent that PopDanceSet, while second only to the AIST++ in terms of dance duration, has comprehensively surpassed it in aspects such as dance and music genres. The collected dance genres encompass 19 categories, including CPOP, KPOP, house dance, among others, and we have endeavored to maintain an even distribution of the number of each dance type. Details of the dataset can be refer to supplementary \cref{sec:popdataset}. The music in this dataset spans a wide range of rhythms and styles, from classical to rock, and most retain lyrics, maintaining consistency with real-world dance environments. The complexity of movements in this dataset also exceeds that of the AIST++, meriting further research in the future.

It is particularly noteworthy that the dance data collected in the PopDanceSet consists of the pose data of human body joints for each frame and the position data in three-dimensional space, without any other parameters such as facial features or body shape. Furthermore, the accompanying music for the dances is all publicly available. Therefore, PopDanceSet does not involve any issues of privacy.

\section{Method}
\label{sec:method}

\begin{figure*}[t]
  \centering
   \includegraphics[width=0.75\textwidth]{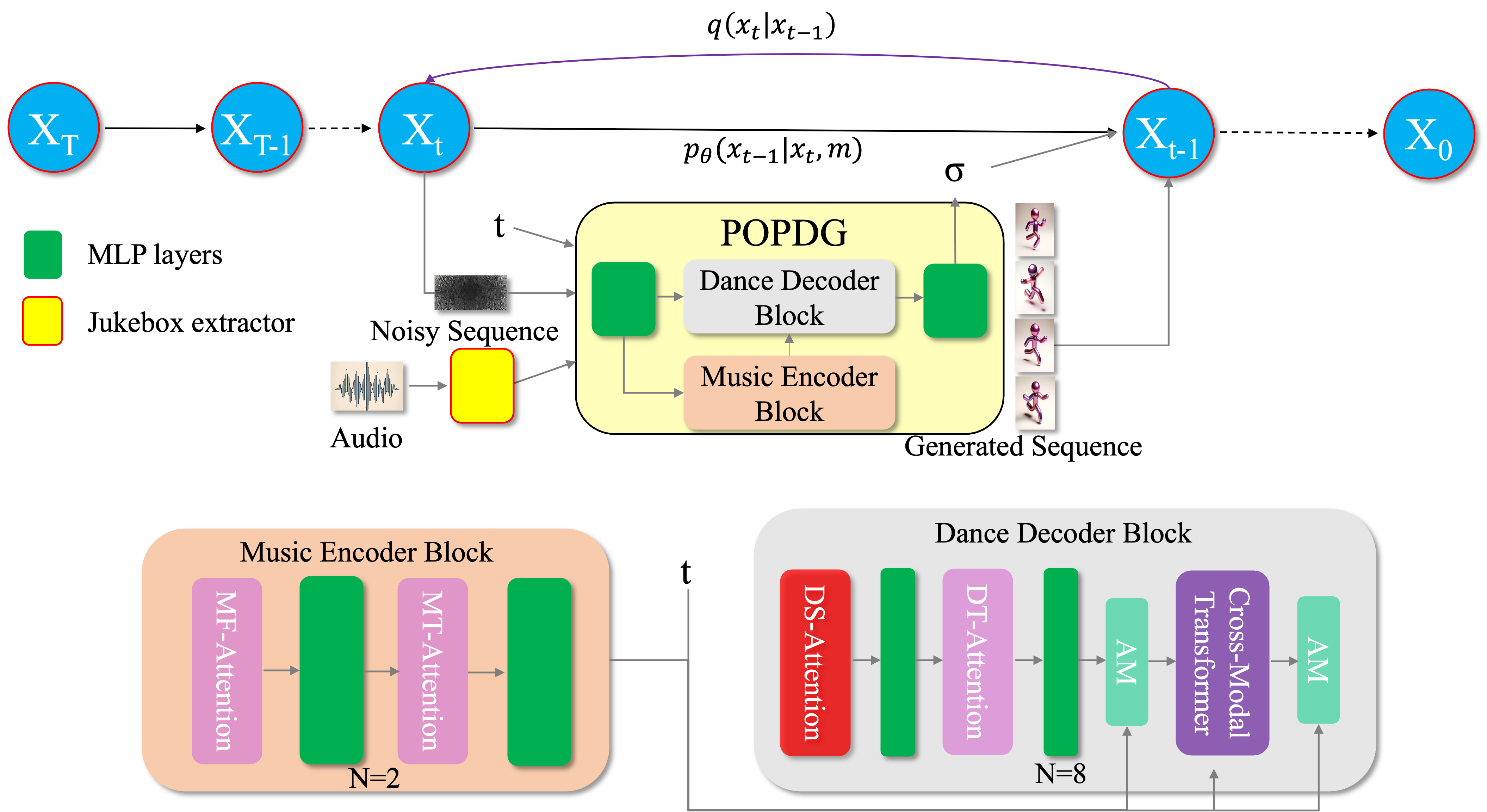}
   \caption{\textbf{POPDG Pipeline Overview.} POPDG, utilizing the iDDPM framework, learns to denoise dance sequences from time \( t = T \) to \( t = 0 \). The audio feature sequence serves as the input to the Music Encoder Block, while the noisy sequence is input to the Dance Decoder Block, with the output being the generated dance sequence. And N refers to the stack number. Beginning with a noisy sequence \( z_{T} \sim N(0,I) \), POPDG generates the estimated frame of the dance sequence. It then progressively noises the sequence back to \( \hat{z}_{T} - 1 \), repeating the process until \( t = 0 \).}
   \label{fig:model}
   \vspace{-2mm}
\end{figure*}
Our model framework, as illustrated in \cref{fig:model}, is based on iDDPM(improved-Denoising Diffusion Probabilistic Models)\cite{nichol2021improved} for sampling with denoising performed by DDIM(Denoising Diffusion Implicit Models)\cite{song2020denoising}. During training, the model is fed a sequence of dance poses \( x \in \mathbb{R}^{N\times156} \)spanning a certain number of frames. In line with most methods, the initial three dimensions represent the single root translation, followed by the 6-DOF (Degrees of Freedom) rotation representation of the 24 joints in the SMPL human body model. The final dimensions are binary contact labels for the feet, hands, and neck. Therefore, the pose representation is \( x \in \mathbb{R}^{156=3+24\cdot6+9} \) per frame. The model gives equal importance to music and motion. In addition to extracting 4800-dimensional music features through Jukebox\cite{dhariwal2020jukebox} like EDGE, it also features a music encoder with a structure symmetrical to that of the dance decoder.


\subsection{Improved-DDPM and DDIM}

Currently, the DDPM framework is employed in both the action domain and dance generation domain, while iDDPM, in comparison to DDPM, learns not only the mean from the data distribution but also takes variance into account, thereby increasing the diversity of generation while ensuring quality. In iDDPM, the forward process also adheres to a Markov chain \( q(z_t|x)\), and we calculate the mean and variance using the following \cref{eq:mean} and \cref{eq:variance}:
\begin{equation}
  q(\boldsymbol{z}_t|\boldsymbol{x})\sim\mathcal{N}(\sqrt{\bar{\alpha}_t}\boldsymbol{x},\boldsymbol{\Sigma_t}),
  \label{eq:mean}
\end{equation}
\begin{equation}
  \Sigma(\boldsymbol{x,t})=\exp(v\log\beta_t+(1-v)\log\tilde{\beta}_t)
  \label{eq:variance}
\end{equation}
where both alpha \(\bar{\alpha}_t \in (0,1) \)and \( \beta_t \) are hyper parameters, and the variable \( v \) is predicted by our model. In reverse process, we learn to estimate \(\boldsymbol{\hat{x}}_\theta(\boldsymbol{z}_t,t,\boldsymbol{m})\approx\boldsymbol{x}\) with model parameter \( \theta \) for all \( t\) . We optimize the basic loss as \cref{eq:loss_simple}:
\begin{equation}
\mathcal{L}_{\mathrm{simple}}=\mathbb{E}_{\boldsymbol{x},t}\left[\|\boldsymbol{x}-\boldsymbol{\hat{x}}_\theta(\boldsymbol{z}_t,t,\boldsymbol{m})\|_2^2\right]
  \label{eq:loss_simple}
\end{equation}
However, this does not take into account variance. Therefore, we follow the iDDPM method, adding a variational lower bound loss. The parameters here also adhere to iDDPM, as shown in \cref{eq:loss_vlb}:
\begin{equation}
  L_{\mathrm{vlb}}=E_{t\sim p_t}\left\lceil\frac{L_t}{p_t}\right\rceil\text{,where }p_t\propto\sqrt{E[L_t^2]}\mathrm{~and~}\sum p_t=1
  \label{eq:loss_vlb}
\end{equation}
Thus, the total loss of iDDPM combines \cref{eq:loss_simple} and \cref{eq:loss_vlb}:
\begin{equation}
  L_{\mathrm{hybrid}}=L_{\mathrm{simple}}+\lambda L_{\mathrm{vlb}}
  \label{eq:loss_hybrid}
\end{equation}
For the denoising process, we follow the DDIM method. This method allows for significantly faster training and inference without much compromise on the quality of the generation.

\subsection{Music and Dance Spatiotemporal block}
As shown in \cref{fig:model}, POPDG comprises two blocks: the music encoder and the dance decoder. These two blocks, based on the principle of symmetry and empirical validation, have similar spatiotemporal Transformer modules. The core of two blocks lies in four attention mechanisms: MF-Attention (Music Feature-Attention), MT-Attention (Music Temporal-Attention), DS-Attention (Dance Spatial-Attention) and DT-Attention (Dance Temporal-Attention).

\subsubsection{Dance Decoder Block}
The details of the dance decoder block can be found in \cref{fig:model}, which is composed of Transformer based on DS-Attention and DT-Attention. Previous methods primarily used DT-Attention. The input dance sequence is 
$x_{motion}$. We take the positional encoded $x_{motion}$ as $Q,K$ and the original $x_{motion}$ as $V$, and pass through the classic attention\cite{vaswani2017attention}:
\begin{equation}
\mathrm{Attention}(\mathbf{Q},\mathbf{K},\mathbf{V},\mathbf{M})=\mathrm{softmax}\left(\frac{\mathbf{Q}\mathbf{K}^T+\mathbf{M}}{\sqrt{C}}\right)\mathbf{V}
  \label{eq:Transformer}
\end{equation}
where $Q*K^{T}$ results in attention map with $[N \times N]$. M represents Mask operation. We randomly mask some frames in the dance sequence to enhance robustness. It mainly pays attention to the temporal relationship within the input sequence.

\begin{figure}[t]
  \centering
   \includegraphics[width=0.65\linewidth]{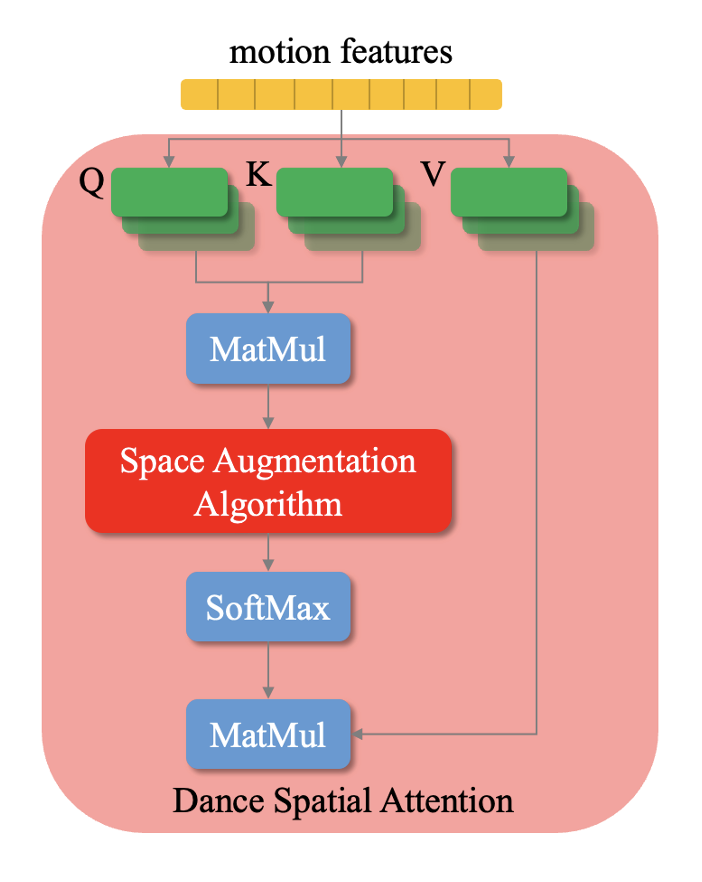}
   \caption{\textbf{The Overview of Dance Spatial Attention.}
   The key distinction between dance spatial attention and standard multi-head attention is the incorporation of the Space Augmentation Algorithm when calculating the Attention Map between Query and Key. This algorithm is tailored to emphasize the upper body joints in relation to the hip, enhancing their spatial inter connectivity.}
   \label{fig:DS-attention}
   \vspace{-3mm}
\end{figure}

In POPDG, we place a spatial attention, DS-Attention, to capture the spatial connections between human body joints, as illustrated in \cref{fig:DS-attention}. In the model, \(x_{motion} \in \mathbb{R}^{b \times N \times h}\), where h represents the hidden feature dimension of the dance posture. We transpose \(x_{motion}\) before feeding it into DS-Attention, thereby obtaining an attention map focused on the spatial dimension. The additional Space Augmentation Algorithm within DS-Attention is capable of capturing the actual spatial connections between the joints.

\begin{figure}[t]
  \centering
    \includegraphics[width=0.95\linewidth]{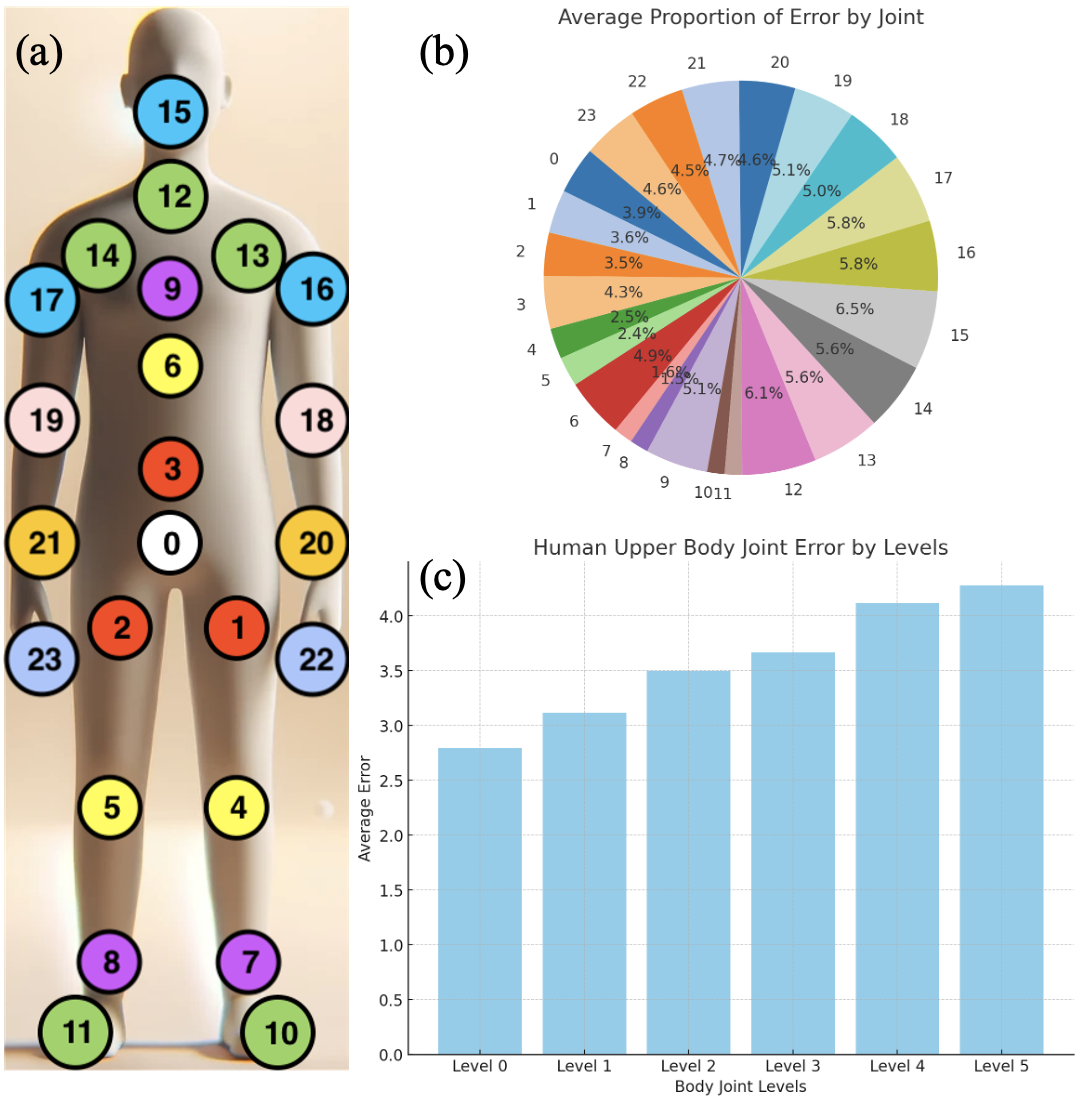}
   \caption{\textbf{Analysis of Joint Error Distribution in SMPL Human Body Model.}
   (a) SMPL Joint Labeling: Marks human body joints from the hip (level $0$ joint) outward, color-coded by different levels. (b) Joint Error Proportions: Shows that upper body joints experience increasing error the further they are from the hip. (c) Upper Body Joint Error Levels: Displays average errors across upper body joint levels.}
   \label{fig:smpl}
\end{figure}

SMPL designates the hip as the root joint, with other body joints categorized into levels based on their distance from the hip. As shown in \cref{fig:smpl}(a), joints with the same background color are at the same level. Comparing generated dance movements with ground truth data, we observe in \cref{fig:smpl}(b) that joint errors increase with distance from the root joint. The average error spans from 4\% at the root to approximately 6.5\% at the upper body parts like the ribs and neck. \cref{fig:smpl}(c) displays the average error across different joint levels.

In the traditional multi-head attention model, attention weights are calculated based on the similarity between queries and keys. To better capture the spatial relationship between specific joints, we introduced a \textbf{Space Augmentation Algorithm}. This algorithm enhances weights based on the distance of the joints from the root joint.

The physical meaning of the algorithm is to strengthen the relationship between each joint in the upper body and its parent joint. Specifically, assuming the $(i+1)^{th} $ joint m is above the $i^{th}$ joint n. It is known that in the calculation of the \(attention map \in \mathbb{R}^{[24,24]} \) in DS-Attention, we already have $(0,m), (m,0)$ and $(m,n),(n,m)$ which represent the connection weights of joint m with the root joint and joint n, respectively. If we add the value of $(0,n)$ to $(0,m)$ and $(m,0)$, essentially we enhance the connection between joint m and the root joint. Similarly, we can continuously pass information from parent joints to downstream-level joints. The specific algorithm implementation can be summarized by \cref{algorithm:algorithm1}. From an experimental perspective, it is viable with or without dividing by 2.

\begin{algorithm}
\caption{\textbf{Space Augmentation Algorithm}}
\begin{algorithmic}[1]
\Function{ApplyWeighting}{$attn\_probs$}
    \State $levels \gets \{ 0: [3], 3: [6], 6: [9], 9: [12, 13, 14], 12: [15], 13: [16], 14: [17] \}$
    \For{$source, targets$ \textbf{in} $levels$}
        \For{$target$ \textbf{in} $targets$}
            \State \Call{Enhance}{$attn\_probs, source, target$}
        \EndFor
    \EndFor
    \State \Return $attn\_probs$
\EndFunction
\Function{Enhance}{$attn, src, tgt$}
    \State $attn[0,tgt] \mathrel{+}= attn[0,src];\ (attn[0,tgt] \mathrel{/}= 2)$
    \State $attn[tgt,0] \mathrel{+}= attn[src,0];\ (attn[tgt,0] \mathrel{/}= 2)$
\EndFunction
\end{algorithmic}
\label{algorithm:algorithm1}
\end{algorithm}

\subsubsection{Music Encoder Block}
Following the principle of symmetry, we adopted a design for the music encoder block that mirrors that of the dance decoder block, and the ablation experiments are displayed in Section 5.4. Building on the existing MT-Attention, we transpose the music data. Unlike dance motions with clear temporal and spatial definitions, after passing through MF-Attention, the musical feature $x_{music}$ will obtain relationships between mathematical features such as MFCC and chroma.

\subsection{Alignment Module}
The quality of generated dance is also contingent on its compatibility with the music. Therefore, building upon the work in Section 4.2, we designed a concise alignment module that can enhance the adaptability of dance to music while ensuring the quality of dance generation.

\begin{figure}[t]
  \centering
   \includegraphics[width=0.65\linewidth]{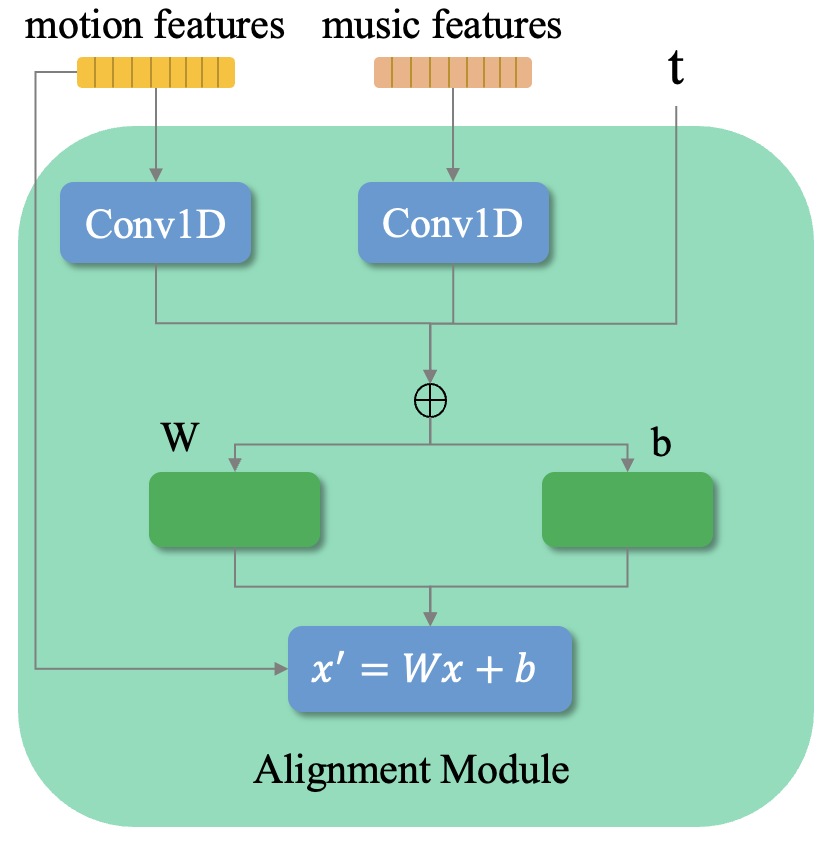}
   \caption{\textbf{The overview of Alignment Module:}
   Once the music and dance features have been processed through temporal and spatial Transformers, we apply temporal feature processing to both.}
   \label{fig:AM}
   \vspace{-4mm}
\end{figure}

Before feeding the dance and music data into our module, we apply temporal processing to both. Unlike previous methods generally applied spatial position encoding to dance sequences, our work equally emphasizes both temporal and spatial characteristics of dance. This involves performing a one-dimensional convolution operation on both sets of features and adding the resultant values to the time step \( t \) in the diffusion model. These combined features are then fed through MLP, consistent with DenseFiLM\cite{radford2021learning}. The detailed model structure is depicted in \cref{fig:AM}.

\subsection{Loss Function}
In the training process, in addition to the initial loss function given by \cref{eq:loss_hybrid}, we integrated and promoted the training strategies of previous methods. Additional information is in supplementary \cref{sec:loss function}:
\begin{itemize}
    \item velocity and acceleration loss:
    \begin{equation}
    \mathcal{L}_{\mathrm{va}}=\frac1{N-1}\sum_{i=1}^{N-1}\|(\boldsymbol{x}^{'}-\boldsymbol{\hat{x}}^{'})\|_2^2+\|(\boldsymbol{x}^{''}-\boldsymbol{\hat{x}}^{''})\|_2^2
    \end{equation}
    We calculate the average error of speed and acceleration between generated dance \( x\) and real dance \( \hat{x} \).
\end{itemize}
\begin{itemize}
    \item FK loss and Body Loss:
    \begin{equation}
    \mathcal{L}_{\mathrm{body}}=\frac1{N-1}\sum_{i=1}^{N-1}\|(FK(\boldsymbol{\hat{x}}^{(i+1)})-FK(\boldsymbol{\hat{x}}^{(i)}))\cdot\boldsymbol{\hat{b}}^{(i)}\|_2^2
      \label{eq:bodyloss}
    \end{equation}
    We adopt the same FK loss function as used in the EDGE. While \(L_{body}\) upgrades \(L_{contact}\), extending its focus from solely the feet to include the hands and the neck. FK(·) denotes the forward kinematic function that converts joint angles into joint positions. \( \hat{b}^{(i)}\) is the model’s own prediction of the binary body contact label’s portion of the pose at each frame i.
\end{itemize}

Finally, by combining these loss functions, we formulate the loss function used for training POPDG, as \cref{eq:wholeloss}.
\begin{equation}
\mathcal{L}=\mathcal{L}_\mathrm{hybrid}+\lambda_\mathrm{FK}\mathcal{L}_\mathrm{FK}+\lambda_\mathrm{va}\mathcal{L}_\mathrm{va}+\lambda_\mathrm{body}\mathcal{L}_\mathrm{body}
      \label{eq:wholeloss}
    \end{equation}
\section{Experiments}
\label{sec:experiments}


\subsection{Implement Details}

In this study, the generative capabilities of the POPDG model are demonstrated on PopDanceSet and AIST++. Initially, to ensure the intrinsic generation quality of the dataset, the construction process included manual checks to confirm the reliability of the extracted dance generation quality. For the PopDanceSet training, there were 736 video segments utilized as the training set and 24 video segments used for testing, ensuring that the dances and music in the test set had not appeared in the training set. The experimental procedure on the AIST++ mirrored the previous methodology, with the training and test sets comprising 952 and 40 videos, respectively, and generating dance sequences lasting 25 seconds in duration.

The entire experimental process took around 100 hours on two A800 GPUs for the AIST++ and 66 hours for the PopDanceSet. The dance decoder block's parameter settings were similar to those used for 3D pose estimation, with the hidden layer dimension uniformly at 512, and MLP layers dimension at 1024. These two parameters were consistently applied in the music encoder block. DT-Attention, MF-Attention, and MT-Attention all employed the conventional 8-head attention mechanism. The optimizer chosen for the model was Adan, with a learning rate set at 0.001 and betas of 0.02, 0.08, and 0.01, with an eps of 1e-8.

\subsection{Evaluation Metrics}

\subsubsection{Motion Quality}
Researchers commonly employ the FID (Frechet Inception Distance) \cite{naeem2020reliable} metric to assess the motion quality of generated dances. However, experiments often reveal that, despite some dances scoring well on FID, they exhibit poor visual quality. In response, EDGE introduced the Physical Foot Contact (PFC) score, denoted by \cref{eq:si}, which assesses the plausibility of dance movements directly through the acceleration of the hips and the velocity of the feet. But since PFC only considers the lower body and dance is a full-body movement, it is also important to consider the upper body\cite{shikanai2014movement}. Therefore, this paper builds upon PFC by including the neck and hands, extending the evaluation to the full body to create the PBC (Physical Body Contact) score.

\begin{equation}
\vspace{-4mm}
    f(x, y, z) = \frac{\sum_{i=1}^{N}||\overline{\boldsymbol{a}}_{x}^i|| \cdot ||\mathbf{v}_{y}^i|| \cdot ||\mathbf{v}_{z}^i||}{\max_{1 \leq j \leq N} ||\overline{a}_{x}^j||}
    \label{eq:si}
\end{equation}
\begin{align}
PBC & = \; \frac{1}{N} \left[ -f(\text{root}, \text{lfoot}, \text{rfoot}) + f(\text{lchest}, \text{lhand}, \text{null}) \right. \nonumber \\
& \left. + f(\text{rchest}, \text{rhand}, \text{null}) + f(\text{neck}, \text{head}, \text{null}) \right]
\label{eq:pbc}
\end{align}


In \cref{eq:pbc}, the variables \(||\overline{\boldsymbol{a}}_{\mathrm{root}}^j||\), \(||\overline{\boldsymbol{a}}_{\mathrm{lchest}}^j||\), \(||\overline{\boldsymbol{a}}_{\mathrm{rchest}}^j||\) and \(||\overline{\boldsymbol{a}}_{\mathrm{neck}}^j||\) each represent the average acceleration of the root joint, the left and right chest joints, and the neck joint of the SMPL model, respectively, projected onto the XYZ plane for each frame \(i\). Compared to PFC, PBC incorporates a broader consideration of the plausibility of dance movements. For detailed elaboration, please refer to supplementary \cref{sec:PBC}.

\subsubsection{Motion Diversity}

In terms of the diversity of generated dances, we have adopted the $Div_{k}$ and $Div_{g}$ metrics used by previous methods, which measure the average kinematic and geometric distance between generated dances and the ground truth to quantify diversity.


\subsubsection{Motion-Music Correlation}
The match between music and dance is also a crucial factor affecting the quality of generated dances. We also employ the formula from previous models to measure the synchrony between dance and music. The Beat Alignment Score adopted follows FACT, defined as:
\begin{equation}
BeatAlign = \frac1{|B^m|}\sum_{t^m\in B^m}\exp\left\{-\frac{\min_{t^d\in B^d}\|t^d-t^m\|^2}{2\sigma^2}\right\}
    \label{eq:BAS}
\end{equation}
where $B^m$ and $B^d$ record the time of beats in dance and music, respectively. And $\sigma$ is normalized parameter which is set to be 3 in our experiment.

\subsection{Comparing to Existing Methods}

\begin{table*}
  \centering
  \begin{tabular}{@{}l|lcccccc@{}}
    \toprule
    & & \multicolumn{2}{c}{Motion Quality} & \multicolumn{2}{c}{Motion Diversity} & \multicolumn{2}{c}{Motion-Music Corr} \\
    \cmidrule(lr){3-4} \cmidrule(lr){5-6} \cmidrule(lr){7-8}
    Dataset & \multicolumn{1}{c|}{Method} & PFC \(\downarrow\) & PBC \(\rightarrow\) & Div\textsubscript{k} \(\uparrow\) & Div\textsubscript{g} \(\uparrow\) & Beat Align Scores \(\uparrow\) \\
    \midrule
    \multirow{5}{*}{PopDanceSet} & GroundTruth  & 1.2302 & 2.8485 & 6.4034 & 7.0289 & 0.330\\
    \cmidrule{2-8}
    & FACT  & 7.5663 & 8.1007 & 3.7371 & \underline{5.7843} & 0.405\\
     & Bailando   & 6.1762 & 5.9237 & 4.2253 & 5.5396 & 0.480 \\
     & EDGE   & 5.9701 & 5.8535 & 3.6065 & 5.7350 & 0.475 \\
     & \textbf{POPDG}  & \textbf{4.3697} & \textbf{5.3863} & \textbf{4.8641} & \textbf{6.0228} & \textbf{0.482}\\
    \midrule
    \multirow{5}{*}{AIST++} & Ground Truth  & 0.3152 & 3.6231 & 8.6916 & 7.5159 & 0.510\\
    \cmidrule{2-8}
    & FACT  & 1.1722 & 8.6751 & 7.5213 & \underline{6.6993} & 0.422\\
     & Bailando   & 0.9268 & 6.8409 & 6.2411 & \textbf{5.7120} & 0.467 \\
     & EDGE   & 0.9201 & 6.5191 & 6.1040 & 3.1415 & 0.456 \\
     & \textbf{POPDG}  & \textbf{0.8014} & \textbf{6.2419} & \textbf{7.5374} & 3.6707 & \textbf{0.469}\\
    \bottomrule
  \end{tabular}
  \caption{\textbf{Dance Quality Comparison on the PopDanceSet and AIST++ Test Sets.} For PopDanceSet, we repused the Bailando and EDGE models, and specifically developed a PyTorch version of FACT to generate dances matching POPDG in length. On AIST++, we continued using the top three previous models for comparison. POPDG mostly outperformed others in motion quality, diversity, and music-dance alignment on both datasets. Notably, the high \(Div_g\) scores of FACT are skewed by excessive, meaningless swaying in both datasets. In our metrics, \(\uparrow\) signifies that higher values indicate better performance, \(\downarrow\) implies the opposite, and \(\rightarrow\) represents that values closer to the Ground Truth are better.}
  \label{tab:combined_results}
\end{table*}

As illustrated in \cref{tab:combined_results}, a comparison with existing methods reveals that our experiment demonstrates the superiority of the POPDG over previous models on the AIST++ and PopDanceSet. Specifically, in the PopDanceSet experiment, POPDG outperformed all other methods, achieving the most optimal results. In terms of the PFC and PBC metrics, POPDG surpassed EDGE by 1.6004(26.8\%) and 0.4672(7.98\%)  in motion quality, respectively. Moreover, POPDG also excelled in dance generation diversity, as evidenced by its superior performance on the $Div_{k}$ and $Div_{g}$ metrics, where it improved by 1.2576(34.9\%) and 0.2878(5.02\%) compared to EDGE. The enhancement in diversity can be attributed to the iDDPM, as discussed in Section 5.4, which augments dance diversity by predicting the mean and variance of motion data. Furthermore, in the BAS metric, POPDG also surpassed Bailando, which specifically employed reinforcement learning to enhance this aspect. Also demonstrated in Section 5.4, the AM strengthens the alignment between music and dance. On the AIST++, although POPDG did not exhibit as significant an impact as in the PopDanceSet, it still surpassed previous works in most evaluation metrics.


\subsection{Ablation Studies}

\begin{itemize}
    \item 
    \textbf{Modules in POPDG} 
    \cref{tab:decoder_ablation} shows the impact of incorporating MF-Attention, DS-Attention, and AM on the quality and music alignment of generated dances. By strengthening the connections between the dancer's body joints, we've improved the quality of the generated dances. Adding MF-Attention has also enhanced the outcomes, considering the model's symmetry. While the exact relationships between various mathematical features of music are still unclear, we believe that POPDG has captured their deeper correlations. The newly designed AM has achieved positive results in both dance quality and alignment, likely due to our modulation of dance using the temporal features of both music and dance.
\setlength{\tabcolsep}{4pt} 
\begin{table}[]
    \centering
    \resizebox{\columnwidth}{!}{
    \begin{tabular}{@{}lccc|ccc@{}}
    \toprule
baseline &MF-Attn &DS-Attn &AM &\text{PFC} \textdownarrow &\text{PBC} \textrightarrow &\text{BAS} \textuparrow \\ \midrule
 \checkmark &           &            &    & 0.9600 & 5.5712 &   -   \\
 \checkmark & \checkmark &            &    & 0.9290 & 5.3363 &  -  \\
 \checkmark & \checkmark & \checkmark &    & 0.9116 & 5.0391 &  0.431   \\
 \checkmark & \checkmark & \checkmark & \checkmark & \textbf{0.8487} & \textbf{4.9626} &  \textbf{0.448}  \\ \bottomrule
    \end{tabular}}
    \caption{\textbf{Modules in POPDG Ablation Study Results.} The ablation study, conducted under the constraints of experimental conditions on an NVIDIA 3090 with the model at half dimension, demonstrates the positive impact of each proposed module on dance generation. Significant improvements were observed in PFC, PBC, and BAS metrics as additional modules were integrated.}
    \label{tab:decoder_ablation}
    \vspace{-4mm}
\end{table}

\end{itemize}

\begin{itemize}
    \item
    \textbf{iDDPM}
    The fundamental difference between the DDPM and iDDPM generative frameworks lies in the fact that while DDPM predicts the mean of the generated data, iDDPM takes into account the variance as well. Theoretically, this gives iDDPM a stronger generative capacity compared to DDPM. For the task of music-driven dance generation, there has traditionally been a trade-off between generation quality and diversity, where improvements in the quality of generated dances tend to reduce diversity. However, the use of iDDPM has allowed us to achieve a favorable balance between generation quality and diversity. This is substantiated by the results presented in \cref{tab:test_iDDPM}.
\begin{table}
  \centering
  \begin{tabular}{@{}lcccc@{}}
    \toprule
    Method & \text{PFC} \textdownarrow
    & \text{PBC} \textrightarrow & \text{Div}\textsubscript{k} \textuparrow & \text{Div}\textsubscript{g} \textuparrow \\
    \midrule
    POPDG & \textbf{0.8014} & 6.2419 & 7.5374 & \textbf{3.6707} \\
    \quad w/o iDDPM & 1.1077 & \textbf{6.1515} & \textbf{8.3189} & 3.3699\\
    \bottomrule
  \end{tabular}
  \caption{\textbf{iDDPM Ablation Study Results.}This experiment validates that the iDDPM framework effectively balances motion quality and diversity.}
  \label{tab:test_iDDPM}
  \vspace{-4mm}
\end{table}
\end{itemize}

\subsection{User Study}
\label{sec: userstudy}

\begin{table}
  \centering
  \begin{tabular}{l|c}
    \toprule
            & \multicolumn{1}{c}{User Study} \\
    \cmidrule{2-2}
    Dataset & Our Dataset Wins \\
    \midrule
    PopDanceSet   & - \\
    AIST++  & 70.0\% $\pm$ 20.0\%\\
    \bottomrule
  \end{tabular}
    \caption{\textbf{User Study Results.}The comparative study between PopDanceSet and AIST++ demonstrates a clear preference for our database among modern youth.}
    \label{tab:user_study}
    \vspace{-6mm}
\end{table}

Our user study involved twenty participants. All participants in the user study were between the ages of 23 and 25. We first trained POPDG on PopDanceSet and AIST++, then selected ten segments of wild music as input, and provided the model-generated dance pairings for participants to evaluate. Participants chose which dance segment they found more appealing. As \cref{tab:user_study} indicates, the majority(70\%) preferred the dances from PopDanceSet, demonstrating that PopDanceSet successfully caters to the aesthetic preferences of the young people. The specific visual rendering effects can be referred to in the supplementary \cref{sec:rendering}.

\section{Conclusion and Discussion}
\label{sec:conclusion}
In this study, we introduce PopDanceSet to enrich data in the field, increase the complexity of dance movements, and reflect contemporary aesthetics.
The POPDG model introduced in this paper enhances the connectivity among the dancer's body parts through DS-Attention and improves the alignment between the generated dance and music using AM. Its iDDPM framework maintains a balance between dance quality and diversity, achieving great results on both PopDanceSet and AIST++. While the model is trained end-to-end, the training cost remains relatively high. Future research should explore strategies to balance the diversity and quality of generated dances using more lightweight models. Additionally, developing metrics that can objectively evaluate dance quality is also crucial for the task. And we also believe that 
a deeper study of music deserves our continued effort.
{
    \small
    \bibliographystyle{ieeenat_fullname}
    \bibliography{main}
}

\clearpage
\setcounter{page}{1}
\maketitlesupplementary

\section{PopDanceSet Construction Details}
\label{sec:popdataset}

\begin{figure}[t]
  \centering
    \includegraphics[width=1\linewidth]{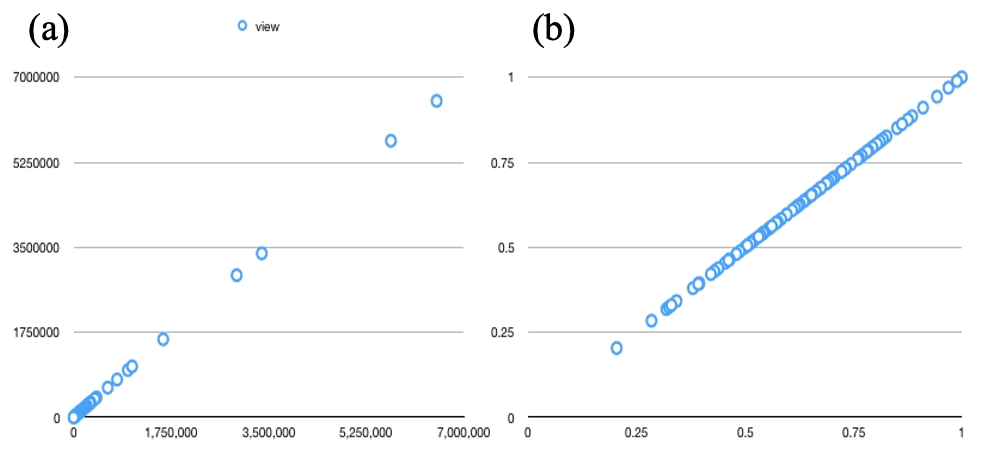}
   \caption{\textbf{Data Preprocessing.}
   (a) The graph represents the original distribution of the view counts for dance videos, showing significant variation in the data distribution; (b) This is the view counts after log normalization, which now exhibits a much more even distribution.}
   \label{fig:data_preprocess}
\end{figure}

The POPDataset was established in September 2022, with the dance videos primarily spanning from September 15, 2021, to September 15, 2022. In the data preprocessing phase, we initially randomly selected 100 dance videos from BiliBili's dance section and collected data on the number of coins, favorites, danmus, comments, views, likes, and shares.  We visualized the view count data as in \cref{fig:data_preprocess} (a).

From \cref{fig:data_preprocess} (a), it is evident that top popular videos have view counts several orders of magnitude higher than average popular videos, and the same happens to other factors. Therefore, direct linear normalization is not suitable in this case. Instead, we employ non-linear normalization (log normalization) for preprocessing the data of the videos, as shown in \cref{fig:data_preprocess} (b).

The core of this experiment in selecting popular dance videos lies in constructing a popularity function. BiliBili's recommendation algorithm for dance videos is given by \cref{eq:recommend}\cite{bilibili}.
\begin{equation}
Recommendation=\frac{W*N}{n_{views}}
  \label{eq:recommend}
\end{equation}
 where \(W=[1.2,0.9,1.2,1.2,0.75,1.2,1.8]\) and \(N=[n_{coins},n_{favorites},n_{danmucounts},n_{comments},n_{views},\\
 n_{likes},n_{shares}]^T\). This formula indicates that the recommendation function considers multiple factors of a video, not just its view count. The function calculates the growth of these facotrs within a specific time frame, with a \(Recommendation\) value greater than 1 significantly increasing the likelihood of the video being recommended on the homepage. Our popularity function was built upon this basis. By omitting the denominator in the formula, we obtained the total values of the video up to the time of data collection. We can then select relevant variables through multiple linear regression and t-tests, with results as in \cref{tab:regression_results}:

\begin{table}[ht]
\centering
\caption{Estimated Value Ranges of Variables from Multiple Linear Regression and T-Test}
\label{tab:regression_results}
\begin{tabular}{lcc}
\toprule
Variable & Lower Bound & Upper Bound \\
\midrule
bias & 0.042 & 0.046 \\
\( n_{\text{coins}} \) & -0.002 & 0.008 \\
\( n_{\text{favorites}} \) & 0.019 & 0.030 \\
\( n_{\text{danmucounts}} \) & 0.004 & 0.011 \\
\( n_{\text{comments}} \) & -0.002 & 0.008 \\
\( n_{\text{views}} \) & 0.798 & 0.814 \\
\( n_{\text{likes}} \) & 0.086 & 0.098 \\
\( n_{\text{shares}} \) & 0.019 & 0.027 \\
\bottomrule
\end{tabular}
\caption*{Note: This table presents the lower and upper bounds of variable estimates resulting from a multiple linear regression analysis followed by a T-test. The bounds signify the expected range of values for each variable.}
\end{table}

Thus, we eliminated the number of coins and comments from the model and, after another round of multiple linear regression and t-test, obtain the formula presented as \cref{eq:heat function}. 

Following the selection measures described, we ultimately filtered out 263 (around 10\% of all collected data) dance videos with a POP greater than 0.85 from a year's span of videos. We also edited these videos into 760 clips featuring relatively high-quality dance content, distributed as \cref{tab:dance_duration}:

\begin{table}[ht]
\centering
\caption{Statistical Distribution of Dance Duration}
\label{tab:dance_duration}
\begin{tabular}{|c|c|c|c|}
\hline
\multicolumn{3}{|c|}{Duration (s)} & \multirow{2}{*}{Total (s)} \\
\cline{1-3}
Short & Medium & Long & \\
\hline
352 (46.3\%) & 394 (51.8\%) & 14 (1.8\%) & 12818.924 \\
\cline{1-4}
\end{tabular}
\caption*{Note: Duration categories are defined as Short (\textless 12s), Medium (12-29.5s), and Long (\textgreater 29.5s). Percentages represent the proportion of total dances falling within each category.}
\end{table}

\begin{figure*}[t]
  \centering
   \includegraphics[width=0.8\textwidth]{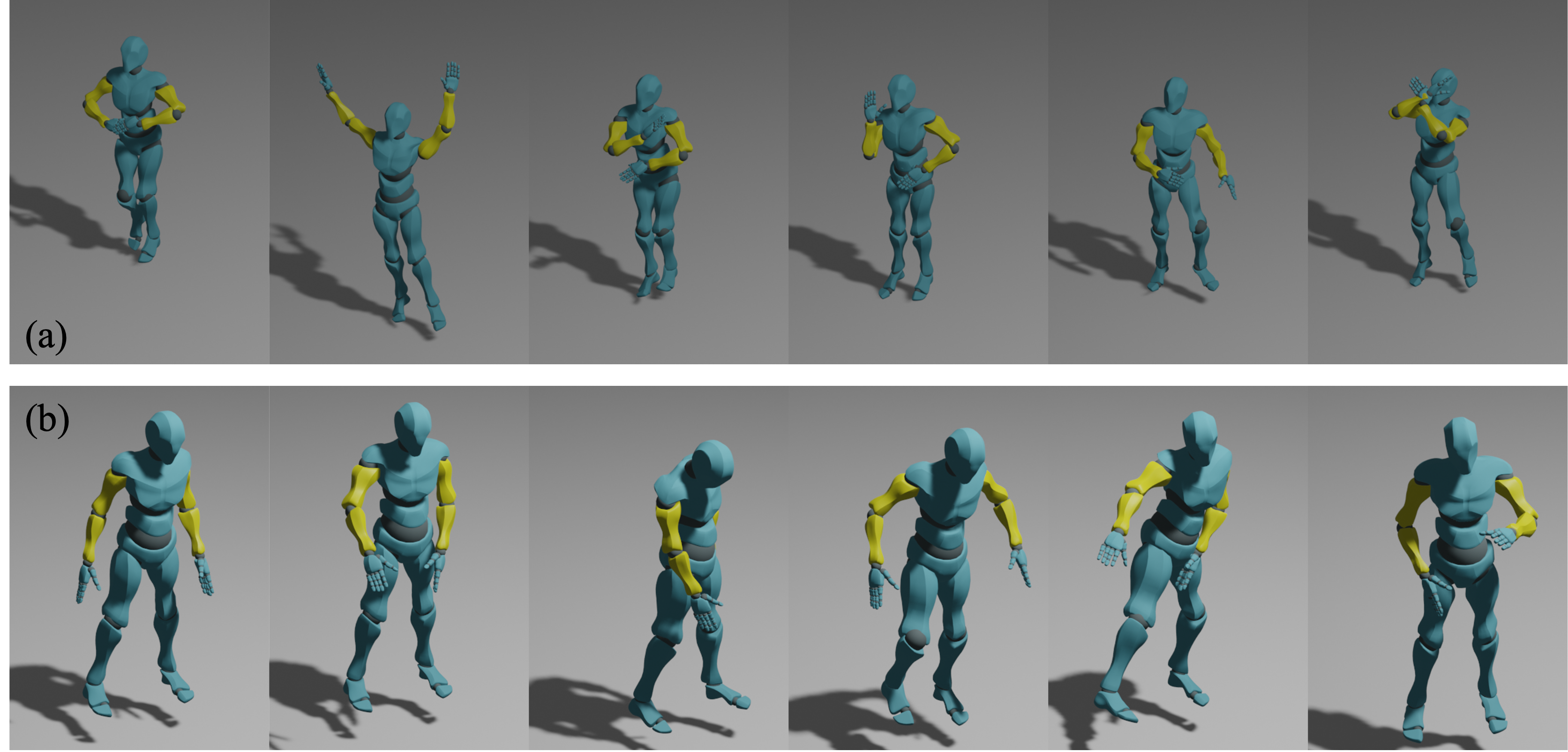}
   \caption{\textbf{Comparison of visual effects between PopDanceSet and AIST++.} (a) shows the dance generation results from PopDanceSet, and (b) shows those from AIST++. Both are comparisons of dance postures at the same frame every second under the same background music. Compared to dances generated based on AIST++, PopDanceSet undoubtedly exhibits richer and more captivating movements.}
   \label{fig:renders}
\end{figure*}

\section{Loss Function}
\label{sec:loss function}
The whole loss is shown as \cref{eq:wholeloss}. And in \cref{sec:method}.4 we have already shown velocity and acceleration loss and body loss, here is the FK loss, as \cref{eq:FKloss}:
\begin{equation}
\mathcal{L}_{\mathrm{FK}}=\frac1N\sum_{i=1}^N\|FK(\boldsymbol{x}^{(i)})-FK(\boldsymbol{\hat{x}}^{(i)})\|_2^2
  \label{eq:FKloss}
\end{equation}
As mentioned in \cref{sec:method}.4, FK(·) denotes the forward kinematic function that converts joint angles into joint positions. Therefore, FK loss is the positional comparison between the generated dance and ground truth in 3D space.

\section{PBC and PFC}
\label{sec:PBC}
The EDGE \cite{tseng2023edge} constructs the PFC (Physical Foot Contact) evaluation metric based on the following two assumptions:
\begin{itemize}
    \item On the horizontal (xy) plane, any center of mass (COM) acceleration must be due to static contact between the feet and the ground. Therefore, either at least one foot is stationary on the ground or the COM is not accelerating.
    \item On the vertical (z) axis, any positive COM acceleration must be due to static foot contact.
\end{itemize}

The PFC derived from these two assumptions is \cref{eq:PFC_origin}:
\begin{equation}
s^i=||\overline{\boldsymbol{a}}_\mathrm{COM}^i||\cdot||\mathbf{v}_\mathrm{Left~Foot}^i||\cdot||\mathbf{v}_\mathrm{Right~Foot}^i||,
  \label{eq:PFCs}
\end{equation}
\begin{equation}
PFC=\frac1{N\cdot\max_{1\leq j\leq N}||\overline{\boldsymbol{a}}_{\mathrm{COM}}^j||}\sum_{i=1}^Ns^i,
  \label{eq:PFC_origin}
\end{equation}
where $\overline{\boldsymbol{a}}_{\mathrm{COM}}^i=\begin{pmatrix}a_{\mathrm{COM},x}^i\\a_{\mathrm{COM},y}^i\\\max(a_{\mathrm{COM},z}^i,0)\end{pmatrix}$.

In the SMPL human body model, the COM (Center of Mass) is represented by the 0th joint at the hip, which is also the root joint in \cref{eq:pbc}. The essence of these two assumptions is that if the body's root joint has acceleration in any direction on the XYZ plane, it means at least one foot must be firmly planted on the ground, as it requires force to initiate movement. Since at least one foot is on the ground, the velocity of that foot should be zero. Thus, the core of PFC is to measure the extent of implausible movements where the body's root joint is accelerating while both feet are moving (i.e., both have velocity). However, this calculation only considers the plausibility of lower body dance movements and overlooks the analysis of upper body movements' plausibility, such as the arms, head and neck. For instance, if the generated dance involves minimal lower body movement but excessive upper body swaying, it would be deemed highly plausible under the PFC metric. Therefore, there's a significant need to also take the upper body into consideration.

In dance, although the upper body movements are relatively independent, we can still observe constraints similar to those between the root joint and the feet within the upper body joints. As illustrated in \cref{eq:pbc}, whether the left and right chest (referred to as the left and right inshoulder in the SMPL model) and neck joints (i.e., joints 12, 13, and 14 in \cref{fig:smpl}(a)) accelerate during a dance largely depends on whether the hands and head are moving, that is, whether they have velocity. Specifically, the movements of the hands and head do not necessarily cause movements in the left and right chest and neck joints, but if the latter do move, it generally indicates that the hands and head have also changed position, thus possessing velocity. Unlike \cref{eq:PFC_origin}, which calculates the irrationality of movements, \cref{eq:pbc} adds a calculation for the rationality of movements. Therefore, in PBC, the value of the original PFC needs to be negated, enabling PBC to reasonably calculate the rationality of full-body dance movements.

\section{Visual Effects Comparison}
\label{sec:rendering}

\cref{fig:renders} showcases a comparison of typical dance clips from PopDanceSet and AIST++. As outlined in \cref{sec: userstudy}, comparing dances generated from the same model trained on different datasets under the background of the same wild music allows for a clearer distinction of which dataset's dances are more appealing. From \cref{fig:renders}, it's evident that dances generated from PopDanceSet are noticeably more engaging. In contrast, dances from AIST++ tend to be more rigid, with several seconds of movement being merely slight adjustments of a single pose. Clearly, the diversity of movements from PopDanceSet, especially in the arm parts, makes these dances more captivating. The only drawback is that the AIST++, with its collection of human keypoints data from nine camera angles, offers somewhat greater stability in the dancer's center of mass compared to PopDanceSet.

\end{document}